\documentclass[prl, twocolumn]{revtex4}
\usepackage{graphicx}
\usepackage{color}
\usepackage{amsmath}
\usepackage{tikz}
\usetikzlibrary{shapes,arrows}

\newcommand{\order}{\mathcal{O}}

\tikzstyle{wide}=[draw, minimum size=2em, text width=7.5em, text centered]
\tikzstyle{narrow}=[draw, minimum size=2em, text width=2em, text centered]
\tikzstyle{int}=[draw, minimum size=2em, text width=4.5em, text centered]

\begin{document}
\title{Co-circulation of infectious diseases on networks}
\author{Joel C. Miller}

\begin{abstract}

We consider multiple diseases spreading in a static Configuration Model network.  We make standard assumptions that infection transmits from neighbor to neighbor at a disease-specific rate and infected individuals recover at a disease-specific rate.  Infection by one disease confers immediate and permanent immunity to infection by any disease.  Under these assumptions, we find a simple, low-dimensional ordinary differential equations model which captures the global dynamics of the infection.  The dynamics depend strongly on initial conditions.  Although we motivate this article with infectious disease, the model may be adapted to the spread of other infectious agents such as competing political beliefs, rumors, or adoption of new technologies if these are influenced by contacts.  As an example, we demonstrate how to model an infectious disease which can be prevented by a behavior change.
\end{abstract}

\maketitle

\paragraph{Introduction}
Networks have captured the attention of many scientists.  One of the primary interests is in understanding how network structure governs the behavior of dynamic processes spreading on networks~\cite{larremore:neuronal,durrett:vote,pastor-satorras:dynamics}.  This is complicated by difficulty in deriving analytic models, limiting our understanding of the dynamics.  Sometimes we can gain insight into long-term behavior without understanding dynamics, but in many cases the intermediate dynamics governs the long-term outcome.  This is particularly significant when competing processes are occurring in the network.  In this article we study the simultaneous spread of two competing diseases in a Configuration Model network.  Although we focus on disease, other competing ``infectious'' processes, such as a change in behavior in response to a disease~\cite{funk:behavior}, spread of beliefs in a voter model~\cite{durrett:vote}, and ``viral marketing'' of competing technologies~\cite{bharathi:competing_tech} have been studied, and the approach introduced here can be adapted to these applications.

In this article we derive a low-dimensional system of equations capturing the dynamics of competing diseases spreading simultaneously in a Configuration Model network.  We apply the model to investigating possible outcomes of co-circulating diseases.  Prior studies have thoroughly analyzed the effect of network structure such as degree distribution~\cite{newman:spread,pastor-satorras:scale-free,boguna:scalefree,may:dynamics} and heterogeneities in infectiousness and/or susceptibility~\cite{miller:heterogeneity,miller:bounds,kenah:second,trapman:analytical} on disease spread.  Recent work gives insight into the role of partnership duration~\cite{volz:dynamic_network,miller:ebcm_overview,miller:ebcm_structure,miller:ebcm_hierarchy}.  Other investigations focus on the role of clustering~\cite{miller:RSIcluster,miller:random_clustered,newman:cluster_alg,gleeson:clustering_effect,melnik:unreasonable,volz:clustered_result,house:insights,trapman:analytical,eames:clustered}, with limited predictive success.

Models of interacting diseases typically neglect network structure~(\emph{e.g.},~\cite{andreasen:cocirculate,cobey:competition} and many others).  Until recently, models of a single disease spreading through a network have relied on approximation~\cite{eames:pair} or been restricted to final size calculations under the assumption of an asymptotically small initial fraction infected~\cite{newman:spread}.  Extending these approaches to competing diseases~\cite{newman:threshold,karrer:competing,funk:interacting} does not allow us to measure the effect of dynamic interactions, and so results are limited to special cases in which these interactions are not important, such as when one disease spreads before the other.

Our method can be easily adapted to more than $2$ diseases and allows for arbitrarily large initial conditions.  We validate the system by comparison with simulation.  Using our equations, we are able to identify the scalings which separate different regimes.  We discuss these regimes and introduce possible generalizations.

\paragraph{The basic model}
We assume that two diseases spread in a Configuration Model network~\cite{newman:structurereview} (also called a Molloy-Reed network~\cite{MolloyReed}) with degree distribution given by $P(k)$.  For disease $1$ transmission along an edge has rate $\beta_1$ and recovery of infected individuals has rate $\gamma_1$.  For disease $2$ the rates are $\beta_2$ and $\gamma_2$.  A node infected by either disease gains immunity to any further infection.

Our approach is similar to that of~\cite{karrer:message} and is based on~\cite{miller:ebcm_overview}.  We will focus our attention on a \emph{test individual} $u$ (described more fully in the appendix and~\cite{miller:final}), a randomly chosen individual in the population.  We assume that the aggregate population-scale spread of the diseases is deterministic.  Under these assumptions, the probability the test individual has a given infection status equals the proportion of the population with that status.  Thus by calculating the probability a test individual has a given status, we immediately know the proportion of the population with that status.

We make one change to the test individual $u$: we prevent it from causing infections.  This keeps the status of its partners independent of one another without affecting its own status, and so it has no effect on our calculations of the proportion of the population in each state.  An alternate argument for why this change has no impact is that we have assumed the dynamics are deterministic, while the timing of when (or even if) $u$ is infected is a random variable.  Thus the infections $u$ would cause cannot have any macroscopic impact on the disease dynamics, and this modification of $u$ has no effect.

We take $t=t_0$ to be our ``initial time''.  In practice this may correspond to the time of introduction of a disease if enough individuals are initially infected, or it corresponds to a later time at which enough infection is present that the spread is deterministic.  There are some restrictions on how the initial infections can be distributed, discussed in the appendix.  We choose a test individual $u$ randomly from the population (it may have any status).  We let $v$ be a random neighbor of a random test individual $u$ which had not transmitted to $u$ by $t_0$.  We define $\theta(t)$ to be the probability that at time $t$, $v$ has not transmitted to $u$.  The probability $u$ is susceptible at time $t$ is
\[
S(t) = \psi(\theta(t)) = \sum_k P(k)S(k,t_0)\theta(t)^k
\]
where $S(k,t_0)$ is the probability an individual of degree $k$ is susceptible at $t=t_0$.  We take $I_1$ and $I_2$ [resp $R_1$ and $R_2$]  to be the probabilities that $u$ is infected with [resp has recovered from] the corresponding disease.

To calculate the change of $\theta$, we must know more about the probability $v$ is in any given state.  We define $\phi_S(t)$ to be the probability $v$ is susceptible, $\phi_{I,1}(t)$ and $\phi_{I,2}(t)$ to be the probabilities that $v$ is infected has not transmitted to $u$, and $\phi_R$ to be the probability $v$ is recovered but did not transmit (we do not need to distinguish which disease infected $v$).  Then $\theta=\phi_S+\phi_{I,1}+\phi_{I,2}+\phi_R$.  

We calculate $\phi_S$ similarly to $S$.  If $v$ is initially susceptible we find the probability it has degree $k$ by counting all edges of initially susceptible individuals of degree $k$: $N k P(k) S(k,t_0)$ and dividing by the number of all edges of initially susceptible individuals $\sum_k N k P(k) S(k,t_0)$ ($N$ is population size).  If $v$ has degree $k$ and was initially susceptible, the probability $v$ is still susceptible is $\theta^{k-1}$ (because $u$ is prevented from transmitting to $v$).  This leads to the conclusion that $v$ is susceptible with probability $\phi_S(t) = \phi_S(t_0)\frac{\sum_k k P(k) S(k,t_0) \theta^{k-1}}{\sum_k kP(k) S(k,t_0)} =\phi_S(t_0)\frac{\psi'(\theta)}{\psi'(1)}$.

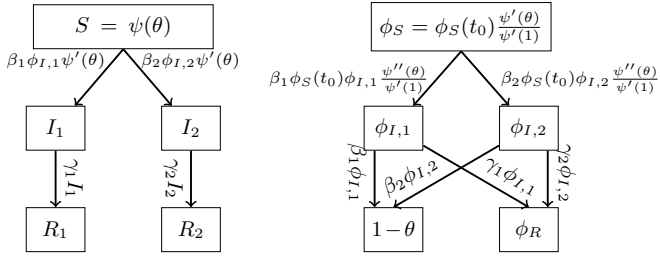
\begin{figure}
\scalebox{0.9}{\begin{tikzpicture}
\node [wide] at (0,0) (S) {$S=\psi(\theta)$};
\node [narrow] at (-1,-1.5) (I1) {$I_1$};
\node [narrow] at (1,-1.5) (I2) {$I_2$};
\node [narrow] at (-1,-3) (R1) {$R_1$};
\node [narrow] at (1,-3) (R2) {$R_2$};
\path [->, thick,left,pos=0.2] (S.270) edge node{{\scriptsize $\beta_1\phi_{I,1}\psi'(\theta)$}} (I1);
\path [->, thick,right,pos=0.2] (S.270) edge node{{\scriptsize $\beta_2\phi_{I,2}\psi'(\theta)$}} (I2);
\path [->, thick,above,sloped] (I1) edge node{{$\gamma_1 I_1$}} (R1);
\path [->, thick,below,sloped] (I2) edge node{{$\gamma_2 I_2$}} (R2);
\end{tikzpicture}
\begin{tikzpicture}
\node [wide] at (0,0) (phiS) {$\phi_S=\phi_S(t_0)\frac{\psi'(\theta)}{\psi'(1)}$};
\node [narrow] at (-1,-1.5) (phiI1) {$\phi_{I,1}$};
\node [narrow] at (1,-1.5) (phiI2) {$\phi_{I,2}$};
\node [narrow] at (1,-3) (phiR) {$\phi_R$};
\node [narrow] at (-1,-3) (1mT) {$1-\theta$};
\path[->, thick, left,pos=0.5] (phiS.270) edge node {{ \scriptsize $\beta_1\phi_S(t_0)\phi_{I,1}\frac{\psi''(\theta)}{\psi'(1)}$
}} (phiI1);
\path[->, thick,right,pos=0.5] (phiS.270) edge node {{ \scriptsize $\beta_2\phi_S(t_0)\phi_{I,2}\frac{\psi''(\theta)}{\psi'(1)}$
}} (phiI2);
\path[->, thick,sloped,above,pos=0.75] (phiI2) edge node {{$\beta_2\phi_{I,2}$}} (1mT.90);
\path[->, thick,sloped,above,pos=0.75] (phiI1) edge node {{$\gamma_1\phi_{I,1}$}} (phiR.90);
\path[->, thick,pos=0.8,below,sloped,pos=0.4] (phiI1.230) edge node {{$\beta_1\phi_{I,1}$}} (1mT.130);
\path[->, thick,pos=0.8,above,sloped,pos = 0.4] (phiI2.310) edge node {{$\gamma_2\phi_{I,2}$}} (phiR.50);
\end{tikzpicture}}
\caption{(left) flow diagram for the probabilities the test individual has each status.  (right) flow diagram for the probabilities a partner of the test individual has each status.}
\label{fig:flow}
\end{figure}

Fig.~\ref{fig:flow} gives flow diagrams which yield our equations.  Each box represents a compartment, and arrow labels represents probability flux from one compartment to another.  The fluxes from the $I$ compartments to the $R$ compartments represent recovery of $u$.  The fluxes from the $\phi_I$ compartments to the $\phi_R$ compartments [resp $1-\theta$] represent flux due to recovery of $v$ prior to transmitting [resp transmission prior to recovery].  The fluxes from $S$ and $\phi_S$ are found by differentiation of $S$ and $\phi_S$ in time, using $\dot{\theta}=-\beta_1\phi_{I,1} - \beta_2\phi_{I,2}$, and assigning the appropriate proportion to the appropriate compartment.

%We now have the pieces needed to create the flow diagrams in Fig.~\ref{fig:flow}.  We begin by looking at compartments for $S$, $I_1$, $I_2$, and $R$, the probabilities $u$ is in each state.  The value of $S$ is $\psi(\theta)$.  The flux of probability from $I_1$ to $R_1$ is simply $\gamma_1I_1$, and similarly for $I_2$ and $R_2$.  So $\dot{R}_1 = \gamma_1 I_1$ and $\dot{R}_2 = \gamma_2 I_2$.  To find the flux into $I_1$ and $I_2$ we differentiate $S$, to find that the flux out is $-\dot{\theta}\psi'(\theta)$.  Taking $\dot{\theta}$ (see below) and attributing the appropriate proportion to each compartment gives flux into $I_1$ of $\beta_1\phi_{I,1}\psi'(\theta)$ and similar flux into $I_2$.  Thus $\dot{I}_1= \beta_1\phi_{I,1}\psi'(\theta)-\gamma_1I_1$ and $\dot{I}_2=\beta_2\phi_{I,2}\psi'(\theta)-\gamma_2I_2$.

%To find the status of $v$ we have similar compartments: $\phi_S$, $\phi_{I,1}$, $\phi_{I,2}$, $\phi_R$, and $1-\theta$ (the probability $v$ has transmitted).  We have an expression for $\phi_S$ in terms of $\theta$.  The other variables can be found by calculating the fluxes between compartments.  The flux into $1-\theta$ is simply $\beta_1\phi_{I,1}$ from the first infected compartment and similarly from the second.  So $\dot{\theta} = -\beta_1\phi_{I,1}+\beta_2\phi_{I,2}$.  The flux into $\phi_R$ is found similarly and $\dot{\phi}_R = \gamma_1\phi_{I,1} + \gamma_2\phi_{I,2}$.  The fluxes into $\phi_{I,1}$ and $\phi_{I,2}$ can be found by differentiating $\phi_S$ to get the flux out of $\phi_S$ and then attributing the appropriate part to each compartment.

From the diagram, we find
\begin{align}
\dot{\theta} &= -\beta_1\phi_{I,1} - \beta_2 \phi_{I,2} \label{eqn:twodiseases_theta_dot}\\
\dot{\phi}_{I,m} &= - (\beta_m+\gamma_m)\phi_{I,m} + \beta_m \phi_{I,m}\phi_S(t_0)\frac{\psi''(\theta)}{\psi'(1)}\\
%\dot{\phi}_{I,2} &= - (\beta_2+\gamma_2)\phi_{I,2} + \beta_2 \phi_{I,2}\phi_S(t_0)\frac{\psi''(\theta)}{\psi'(1)}\\
S &= \psi(\theta)\\
\dot{I}_m &=\beta_m \phi_{I,m} \psi'(\theta)- \gamma_m I_m\\
%\dot{I}_2 &=\beta_2 \phi_{I,2} \psi'(\theta)- \gamma_2 I_2\\
\dot{R}_m &= \gamma_m I_m \label{eqn:twodiseases_R2_dot}
%\dot{R}_2 &= \gamma_2 I_2 
\end{align}
The subscript $m$ takes the values $1$ and $2$.  The single-disease, small initial condition limit of these equations has been proved exact~\cite{decreusefond:volz_limit}.  These equations capture the fact that disconnected components are safe from outside introduction.

\paragraph{Sequential Introduction}
As an example, we consider two diseases spreading in a network of $10^6$ individuals with Poisson degree distribution of mean $5$.  For both, $\gamma=1$ but $\beta_1 = 0.5$ and $\beta_2=1$.  In simulations, we introduce disease $1$ into $30$ random individuals at $t=0$, and disease $2$ into $30$ random individuals at $t=5.25$.  

\begin{figure}
\includegraphics[width=0.4\textwidth]{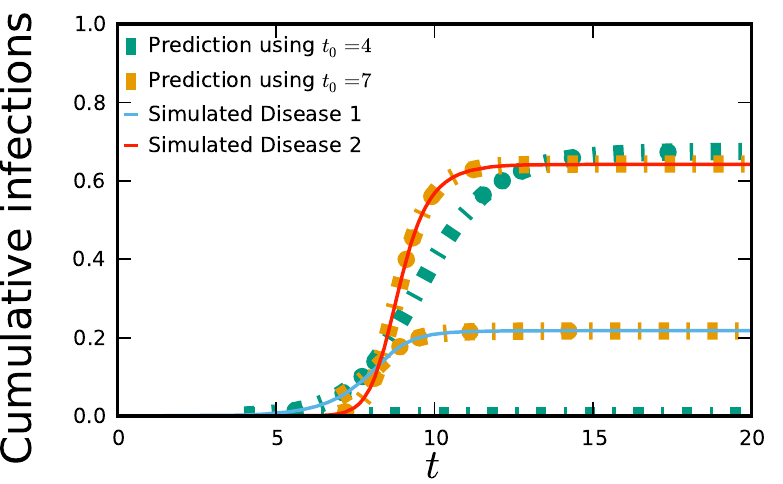}\hfill
\includegraphics[width=0.4\textwidth]{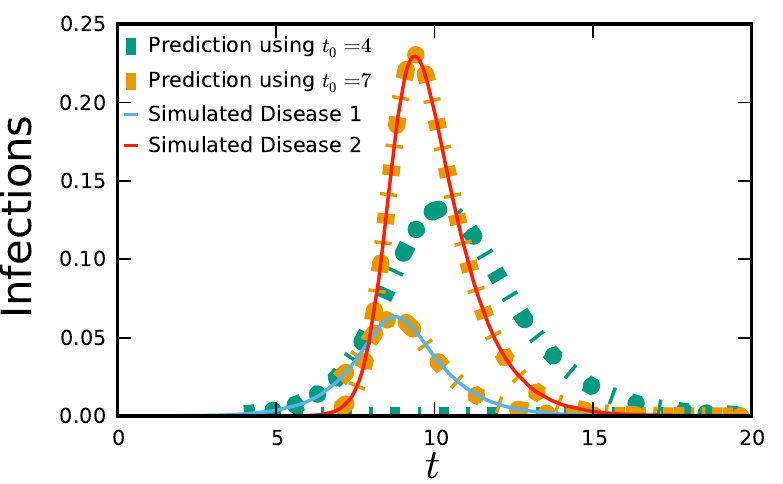}
\caption{The spread of two diseases in a population of size $10^6$ with a Poisson degree distribution of mean $5$.  The first disease is introduced with $30$ cases at $t=0$, and the second with $30$ cases at $t=5.25$.  The second strain is more infectious.  Predictions (dashed) are calculated from observations at $t=4$ (before the second disease's introduction) and $t=7$ (shortly after).  Model and simulation agree well.}
\label{fig:sequential}
\end{figure}

Our deterministic equations do not apply while either disease has a small number of infections.  We take observations at $t_0=4$ (after the first, but before the second disease) and $t_0=7$ (after both are established) to initialize our equations.  Fig.~\ref{fig:sequential} compares calculation with simulation.  Comparing the $t_0=4$ calculation with the $t_0=7$ calculation shows the effect of the second disease.  
%Different simulations have different stochastic phases, and so different dynamic interactions result.

\begin{figure}
\includegraphics[width=0.48\textwidth]{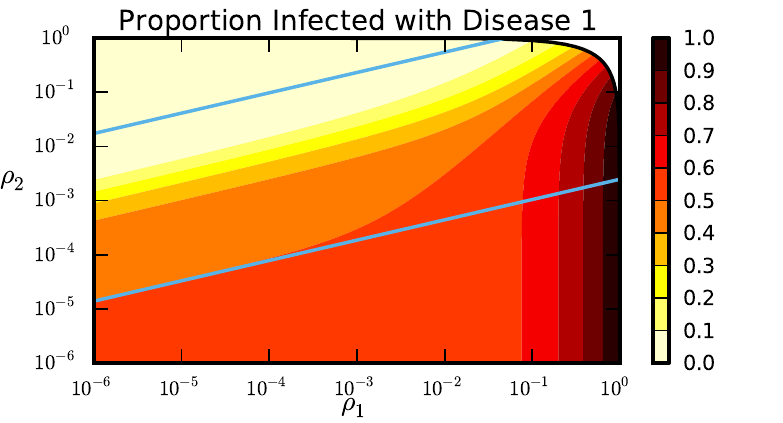}
\includegraphics[width=0.48\textwidth]{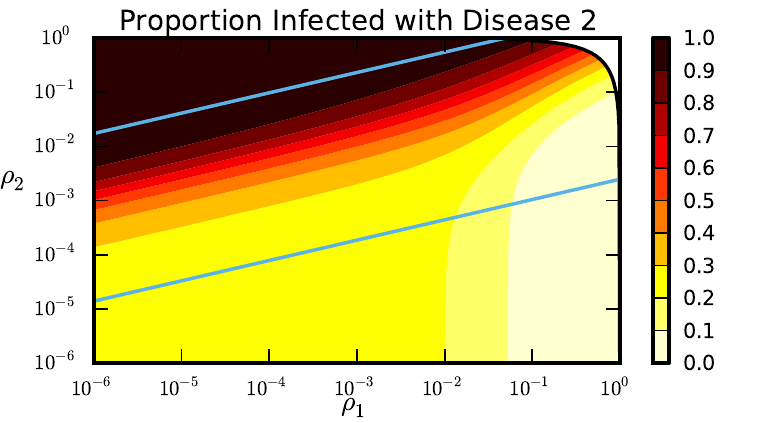}
\caption{  We take a network with the degree of each node chosen uniformly from $1$ up to $9$.  A proportion $\rho_1$ of the population begins infected with disease $1$ and a proportion $\rho_2$ begins infected with disease $2$ with $\beta_1=1.2$, \ $\gamma_1=4$, \ $\beta_2 = 0.2$, and $\gamma_2=0.25$.  Disease $1$ spreads more quickly, but has a smaller per-edge transmission probability.  The solid lines denote the estimated upper and lower bounds of the overlapping epidemic regime.}
\label{fig:final_size_comp}
\end{figure}

\paragraph{Simultaneous Introduction}
We now consider the simultaneous introduction of two diseases and assume that the initial numbers infected are large enough that the dynamics are deterministic.  In our example, we take $\beta_1=1.2$, \ $\gamma_1=4$, \ $\beta_2 = 0.2$, and $\gamma_2=0.25$.  Disease $1$ tends to spread more quickly, but disease $2$ has a higher probability of transmission prior to recovery.  At $t=0$, we infected a randomly selected proportion of the population $\rho_1$ with disease $1$, and a proportion $\rho_2$ with disease $2$.  This gives $S(k,0)=1-\rho_1-\rho_2$ for all $k$, \ $I_1=\rho_1$, \ $I_2 = \rho_2$, \ $\phi_{I,1} = \rho_1$, and $\phi_{I,2}=\rho_2$, with no recovered individuals.  In our population, the degree of each node is assigned uniformly from the integers $1$ through $9$.  We use our equations to calculate the final proportion infected by each disease, shown in Fig.~\ref{fig:final_size_comp}.

There are several distinct regimes we can identify in Fig.~\ref{fig:final_size_comp}.  If $\rho_2$ is $\order(1)$ and $\rho_1$ small, or if $\rho_1$ is $\order(1)$ and $\rho_2$ small, the disease with the large initial condition spreads and effectively infects everyone simply because a large fraction is initially infected.  The other disease cannot spread in the ``residual network'' left behind.  If neither $\rho_1$ or $\rho_2$ is initially large, other regimes are seen.  To analyze them, we first note that when both diseases are small, they grow at exponential rates $r_1 = -(\beta_1+\gamma_1) + \beta_1\psi''(1)/\psi'(1)=2$ and $r_2 = -(\beta_2+\gamma_2) + \beta_2\psi''(1)/\psi'(1)=0.75$.

In the ``overlapping epidemic'' regime, the slower-growing disease $2$ begins with a head start.  The size of the head start scales so that the two epidemics become large at the same time.  The value of $\ln I_2 - (r_2/r_1)\ln I_1$ is constant during the linear growth phase. For given $C = \ln \rho_2 - (r_2/r_1) \ln \rho_1$, the behavior is universal. The diseases grow independently until the linear growth phase ends.  We can estimate bounds on the regime by crudely assuming exponential growth continues forever.  There is some value of $C_{\text{min}} = \ln 0.0025 - (r_2/r_1)\ln 1 = \ln 0.0025 \approx -6$ that corresponds to $\rho_1=1$ and $\rho_2 = 0.0025$ which means that the slower growing disease would affect less than one percent of the population by the time the faster growing disease has fully established itself in this approximation.  Similarly we take some $C_{\text{max}} = \ln 1 - (r_2/r_1)\ln 0.05 = -(r_2/r_1)\ln 0.05\approx 3r_2/r_1$ corresponding to $\rho_1=0.05$ and $\rho_2=1$, which means that the slower growing disease will have fully burned through the population when the faster growing disease is only affecting 5\% of the population in this approximation.  For $C<C_{\text{min}}$, $I_1$ becomes large well before $I_2$.  For $C>C_{\text{max}}$, $I_2$ becomes large well before $I_1$.  Between these values, the epidemics become large at similar times and interact dynamically.  Further discussion of these boundaries is in the appendix.

There are two``non-overlapping epidemic'' regimes.  If $C<C_{\text{min}}$, disease $1$ becomes large and has an epidemic while disease $2$ is still exponentially small.  If $C>C_{\text{max}}$, disease $2$ has an epidemic while disease $1$ is still exponentially small.  Once one epidemic has finished, it may be possible for the remaining disease to spread in the residual network.  We can derive the threshold condition: For simplicity we assume disease $1$ spreads first.  While it spreads, $\phi_{I,2}$ remains negligible.  By looking at the relative fluxes out of $\phi_{I,1}$ we conclude that after disease $1$ has completed its spread but before disease $2$ is significant, $\phi_R = \gamma_1(1-\theta)/\beta_1$.  We can assume $\phi_S(0)=1$ and $\phi_S(t)=\psi'(\theta)/\psi'(1)$.  Since $\phi_{I,1}$ and $\phi_{I,2}$ are effectively zero, we conclude $\theta = \phi_S+\phi_R =  \psi'(\theta)/\psi'(1) + \gamma_1(1-\theta)/\beta_1$, yielding 
\[
\theta = \frac{\beta_1}{\beta_1+\gamma_1}\left(\gamma_1 + \frac{\psi'(\theta)}{\psi'(1)} \right)
\]
Then $R_1 = 1-S(\theta)$.
For disease $2$ to spread, we require $\dot{\phi}_{I,2}>0$, which implies $\beta_2/(\beta_2+\gamma_2) > \psi'(1)/\psi''(\theta)$ where $\theta$ comes from the above equation.  A similar calculation would lead to a final size for $R_2$, so in this case we can calculate the final outcomes of the epidemics without calculating the dynamics.  We note that the threshold we have found requires that $\beta_2/(\beta_2+\gamma_2)$ be greater than $\beta_1/(\beta_1+\gamma_1)$ by a nonzero amount.  It is not enough for the second disease's transmission probability to simply be larger than the first, it must be well above that of the first.  This is because even once disease $1$ peaks and begins to decrease, $\theta$ will continue to decrease further, so disease $2$ encounters a population that is well below the threshold for disease $1$.  The threshold condition for disease $2$ to invade has been identified previously: \cite{newman:threshold} derived it under the assumption of a second disease introduced after the first disease had spread, and \cite{karrer:competing} derived it in the special case that the system was in the non-overlapping epidemic regime for which the faster growing disease would ``win.'' 

Thus our analysis shows that if two diseases are introduced at small levels into the network, then the possible regimes can be understood by looking at the exponential growth rates $r_1$ and $r_2$ of the diseases.  Without loss of generality, we can assume $r_1\geq r_2$, the first disease grows faster.  If disease $2$ has a sufficiently large head start, there will be a ``non-overlapping epidemic'' regime: the population will experience an epidemic of disease $2$ unaffected by disease $1$.  If the infectiousness of disease $1$ is large enough it will cause its own epidemic after disease $2$ has finished.  We can calculate the final size of each epidemic without requiring the full dynamic calculation.  If we take a smaller head start for disease $2$, there is an overlapping epidemic regime in which the two diseases produce interacting epidemics.  To calculate the dynamics of these epidemics, we require the dynamic equations~\eqref{eqn:twodiseases_theta_dot}--\eqref{eqn:twodiseases_R2_dot}.  The possible final sizes depend on the details of the interactions, and there appears to be no simple expression for the final size.  If the head start for disease $2$ shrinks further, we enter another non-overlapping epidemic regime in which disease $1$ has the first epidemic.  Again, it is possible for disease $2$ to later have an epidemic if its transmission probability is sufficiently larger than that of disease $1$.  At most one of the non-overlapping epidemic regimes can have epidemics for both diseases.

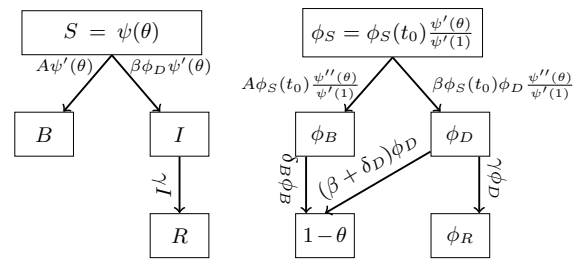
\begin{figure}
% \begin{tikzpicture}
% \node [wide] at (0,0) (phiS) {$\phi_S=\phi_S(0)\frac{\psi'(\theta)}{\psi'(1)}$};
% \node [narrow] at (4,0.75) (phiD) {$\phi_D$};
% \node [narrow] at (4,-0.75) (phiB) {$\phi_B$};
% \node [narrow] at (7,0.75) (phiR) {$\phi_R$};
% \node [narrow] at (7,-0.75) (1mT) {$1-\theta$};
% \path[->, thick, left,pos=0.95] (phiS.0) edge node {{$\beta\phi_D\phi_S(0)\frac{\psi''(\theta)}{\psi'(1)}$}} (phiD.180);
% \path[->, thick, left,pos=0.95, font = \footnotesize] (phiS.0) edge node {{$(\delta_D\phi_D+\delta_B\phi_B)\phi_S(0)\frac{\psi''(\theta)}{\psi'(1)}$}} (phiB.180);
% \path[->, thick,above,sloped] (phiD.340) edge node {{$(\beta+\delta_D)\phi_D$}} (1mT.90);
% \path[->, thick, above] (phiB) edge node {{$\delta_B\phi_B$}} (1mT);
% \path[->, thick,above] (phiD) edge node {{$\gamma\phi_D$}} (phiR);
% %\path[->, thick, right] (phiI2) edge node {{$\gamma_2\phi_{I,2}$}} (phiR);
% \end{tikzpicture}\\[20pt]
% \begin{tikzpicture}
% \node [wide] at (0,0) (S) {$S=\psi(\theta)$};
% \node [narrow] at (4,0.75) (I) {$I$};
% \node [narrow] at (4,-0.75) (B) {$B$};
% \node [narrow] at (7,0.75) (R) {$R$};
% \path [->, thick,left,pos=0.8] (S.0) edge node{{$\beta\phi_D\psi'(\theta)$}} (I.180);
% \path [->, thick,left,pos=0.8] (S.0) edge node{{$(\delta_D\phi_D+\delta_B\phi_B)\psi'(\theta)$}} (B.180);
% \path [->, thick,below] (I) edge node{{$\gamma I$}} (R);
%\end{tikzpicture}
\scalebox{0.9}{\begin{tikzpicture}
\node [wide] at (0,0) (S) {$S=\psi(\theta)$};
\node [narrow] at (-1,-1.5) (B) {$B$};
\node [narrow] at (1,-1.5) (I) {$I$};
\node [narrow] at (1,-3) (R) {$R$};
\path [->, thick,left,pos=0.2] (S.270) edge node{{\scriptsize $A\psi'(\theta)$}} (B);
\path [->, thick,right,pos=0.2] (S.270) edge node{{\scriptsize $\beta\phi_D\psi'(\theta)$}} (I);
\path [->, thick,below,sloped] (I) edge node{{$\gamma I$}} (R);
\end{tikzpicture}
\begin{tikzpicture}
\node [wide] at (0,0) (phiS) {$\phi_S=\phi_S(t_0)\frac{\psi'(\theta)}{\psi'(1)}$};
\node [narrow] at (-1,-1.5) (phiB) {$\phi_B$};
\node [narrow] at (1,-1.5) (phiD) {$\phi_D$};
\node [narrow] at (1,-3) (phiR) {$\phi_R$};
\node [narrow] at (-1,-3) (1mT) {$1-\theta$};
\path[->, thick, left,pos=0.5] (phiS.270) edge node {{ \scriptsize $A\phi_S(t_0)\frac{\psi''(\theta)}{\psi'(1)}$
}} (phiB);
\path[->, thick,right,pos=0.5] (phiS.270) edge node {{ \scriptsize $\beta\phi_S(t_0)\phi_D\frac{\psi''(\theta)}{\psi'(1)}$
}} (phiD);
\path[->, thick,sloped,above,pos=0.5] (phiD) edge node {{$(\beta+\delta_D)\phi_D$}} (1mT.90);
\path[->, thick,pos=0.8,below,sloped,pos=0.4] (phiB.230) edge node {{$\delta_B\phi_B$}} (1mT.130);
\path[->, thick,pos=0.8,above,sloped,pos = 0.4] (phiD.310) edge node {{$\gamma\phi_D$}} (phiR.50);
\end{tikzpicture}}
\caption{The flow diagram for disease spread with behavior change.  Here $A = (\delta_B\phi_B+\delta_D\phi_D)$.}
\label{fig:behaviorflow}
\end{figure}

\paragraph{Generalizations}
This model may be adapted to other ``infectious'' agents such as the spread of a rumor or a new technology through a social network.  As an example of its flexibility we consider a disease which can be avoided through behavior modification.  We assume contact with an infected individual transmits infection at rate $\beta$.  However, we allow that if $u$ is in contact with an infected individual, then at rate $\delta_D$ $u$ changes its behavior.  If $u$ is in contact with an individual who has changed behavior then $u$ changes behavior at rate $\delta_B$.  This leads to the flow diagram in Fig.~\ref{fig:behaviorflow}.

In Fig.~\ref{fig:behave}, we show how behavior change modifies epidemic outcomes.  If the disease is introduced in a small number of individuals and behavior change spreads sufficiently faster than the disease, then the behavior change prevents the disease from having a large-scale epidemic.

\begin{figure}
\includegraphics[width=0.48\textwidth]{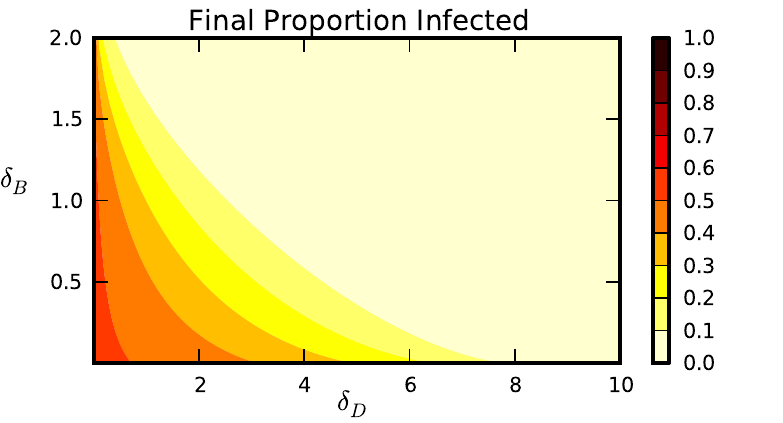}
\caption{Final sizes of epidemics with behavior change.  We use networks of the same structure as in Fig.~\ref{fig:final_size_comp}.  The disease spreads with $\beta = 2$, \ $\gamma=1$.  The values of $\delta_B$ and $\delta_D$ are varied.  For the initial condition, no individuals have changed behavior and a proportion $10^{-6}$ is initially infected randomly.}
\label{fig:behave}
\end{figure}

\paragraph{Summary}
We have introduced an analytic model that calculates the simultaneous spread of two infectious diseases in a configuration model network.  Our model is low-dimensional regardless of the degree distribution.  Using this model, we are able to calculate the effect of interactions between diseases in regimes that are inaccessible to analytic theories that do not include dynamics.

This model can easily be generalized to a range of other ``infectious'' processes. We have shown its application to behavior changes in response to a disease.

\paragraph{Acknowledgments}
This work was supported by 1) the RAPIDD program of the Science and Technology Directorate, Department of Homeland Security and the Fogarty International Center, National Institutes of Health (NIH) and 2) the Center for Communicable Disease Dynamics, Department of Epidemiology, Harvard School of Public Health under Award Number U54GM088558 from the National Institute Of General Medical Sciences (NIGMS).  The content is solely the responsibility of the author and does not necessarily represent the official views of the NIGMS or the NIH.

\appendix
\section{Appendix}

\paragraph{Detailed derivation of equations}
We begin with a configuration model network, and consider two infectious
diseases which infect nodes of the network.  We assume infection by
either disease confers immediate and complete protection from any future
infection by any disease.  We index the diseases by $1$ and $2$.
Individuals infected by disease $i=1,2$ transmit to their partners at
rate $\beta_i$, and recover at rate $\gamma_i$.

We assume that at the initial time $t_0$ enough individuals are
infected that the disease spreads deterministically (at the aggregated
population-level), and that the
probability an individual of degree $k$ is initially infected is
$S(k,t_0)$.  We must make an assumption about which individuals are
initially infected.  Namely, if we consider an initially susceptible
individual $u$, no information we have about $u$ at time $t_0$ tells
us anything about the status of its partners.  This assumption is
satisfied if we initially infect a random subset of the population, or
even if the disease has been spreading for some time before $t_0$.
This assumption is violated if we select high-degree individuals and
preferentially infect their partners.

\paragraph{The test individual}
We now introduce the concept of a \emph{test individual}.  This
concept is described in more detail in~\cite{miller:final}, and it
allows us to simplify our calculations.

Whether a given individual $u$ is infected at any given time is a random
variable.  However, if we make the assumption that the disease spreads
deterministically at the aggregated population-level, then whether or
not a given individual is infected at any given time cannot have any
impact on the aggregated scale.  If it did, there would be stochastic
effects visible at the population scale.

This observation allows us to decouple the status of $u$ from the
dynamics of the epidemic in the sense that we can ignore any feedback
from $u$ on the epidemic.  To make this mathematically rigorous, we
simply allow $u$ to become infected and for its infection to proceed
as normal, but we disallow any transmission from $u$ to its partners.
This keeps the status of partners of $u$ independent.

The probability that $u$ is susceptible equals the probability that
none of its partners has transmitted to it (under the assumption $u$
does not transmit to its partners).  To calculate the proportion of
the population that is susceptible, infected, or recovered, we assume
that $u$ is randomly selected from the population and prevented from
infecting its partners.  We call $u$ a \emph{test individual}.  The
probability $u$ has a given status equals the proportion of the
population that has that status.

\paragraph{Deriving the flow diagrams}
We define $\theta(t)$ to be the probability that a random neighbor of
$u$ which had not transmitted to $u$ by time $t=t_0$ still has not
transmitted by time $t$.  Then if $u$ has degree $k$, the probability
it was initially susceptible is $S(k,t_0)$, and the probability it is
still susceptible is $S(k,t_0)\theta(t)^k$.  Averaging over all
possible values of $k$, we have
\[
S(t) = \psi(\theta(t)) = \sum_k P(k) S(k,t_0)\theta(t)^k
\]
The value of $S$ reduces over time as infections occur.
Mathematically this appears as a reduction in $\theta$.  We must
calculate how quickly $\theta$ changes, and how much of that change is
due to each disease.  This will allow us to calculate how much of the
reduction in $S$ should go into each disease's infected class.

\begin{figure}
\begin{tikzpicture}
\node [wide] at (0,0) (phiS) {$\phi_S=\phi_S(t_0)\frac{\psi'(\theta)}{\psi'(1)}$};
\node [int] at (3,1.5) (phiI1) {$\phi_{I,1}$};
\node [int] at (3,-1.5) (phiI2) {$\phi_{I,2}$};
\node [int] at (6,1.5) (phiR) {$\phi_R$};
\node [int] at (6,-1.5) (1mT) {$1-\theta$};
\path[->, thick, left,pos=0.95] (phiS.0) edge node {{ $\beta_1\phi_S(t_0)\phi_{I,1}\frac{\psi''(\theta)}{\psi'(1)}$~~
}} (phiI1.180);
\path[->, thick,left,pos=0.95] (phiS.0) edge node {{ $\beta_2\phi_S(t_0)\phi_{I,2}\frac{\psi''(\theta)}{\psi'(1)}$~~
}} (phiI2.180);
\path[->, thick,below] (phiI2) edge node {{$\beta_2\phi_{I,2}$}} (1mT);
\path[->, thick,above] (phiI1) edge node {{$\gamma_1\phi_{I,1}$}} (phiR);
\path[->, thick, above,pos=0.8,sloped,pos=0.3] (phiI1.340) edge node {{$\beta_1\phi_{I,1}$}} (1mT.90);
\path[->, thick,above,pos=0.8,sloped,pos = 0.2] (phiI2.20) edge node {{$\gamma_2\phi_{I,2}$}} (phiR.270);
\end{tikzpicture}
\caption{Flow diagram that leads to the evolution of $\theta$.  A
  label along an edge gives the flux of probability along that edge.}
\label{fig:theta}
\end{figure}
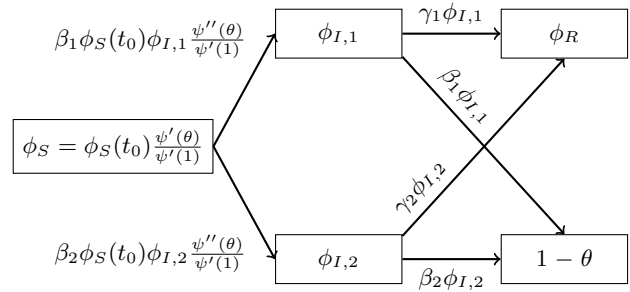

We now look for the change in $\theta$.  We define $v$ to be a random
neighbor of $u$ which had not transmitted to $u$ by $t_0$.  Then
$\theta$ is the probability $v$ has not transmitted to $u$ by time
$t$.  As our initial condition, we have $\theta(t_0)=1$.  We divide
$\theta$ into four compartments.  We take $\phi_S$ to be the
probability that $v$ is still susceptible, $\phi_{I,1}$ the
probability $v$ is infected with disease $1$ but has not transmitted
to $u$, $\phi_{I,2}$ the probability $v$ is infected with disease $2$
but has not transmitted to $u$, and $\phi_R$ the probability that $v$
is recovered (from either disease) and did not transmit during
infection.  These add up to $\theta$: $\theta = \phi_S + \phi_{I,1} +
\phi_{I,2} + \phi_R$.  The probability that $v$ has transmitted to $u$
is $1-\theta$.  The flow between these compartments is shown in
figure~\ref{fig:theta}.

We first find the rate of change of $\theta$.  It is relatively
straightforward to see that 
\[
\dot{\theta} = -\beta_1 \phi_{I,1} -
\beta_2 \phi_{I,2}
\]
This is because the only path for $\theta$ to decrease is through $v$
transmitting to $u$, which requires that $v$ be infected (and not yet
transmitted to $u$).

At $t_0$, we have $\phi_S(t_0)$ is the probability a random neighbor
of $u$ is still susceptible (given that it has not transmitted to
$u$).  We take this as an input value.  The value of $\phi_{I,1}$,
$\phi_{I,2}$, and $\phi_R$ are similarly all input from the conditions
at $t_0$.  We can explicitly calculate the value of $\phi_S$ at later
times, if we know $\theta$.  To do this, we find the probability
distribution for degree of $v$, and then calculate the probability
that no neighbor of $v$ has transmitted to $v$.  

At time $t_0$, the edge joining $u$ to $v$ is simply an edge from $u$
to a random neighbor that is susceptible.  The probability that edge
connects to a degree $k$ individual is proportional to the number of
edges that all susceptible individuals of degree $k$ have, $NkP(k)
S(k,t_0)$.  The normalization factor is the total number of all edges
of susceptible individuals, $\sum_{k'} N k' P(k')S(k',t_0)$.  The
probability that $v$ is still susceptible at time $t$ is
$\theta(t)^{k-1}$.  So the probability of having a degree $k$
susceptible neighbor at time $t$ is $N k P(k) S(k,t_0) \theta(t)^{k-1}
/ \sum_{k'} N k' P(k') S(k',t_0)$.  Summing over all possible $k$, and
cancelling $N$, we arrive at $\sum_k k P(k) S(k,t_0) \theta(t)^{k-1}/
\sum_{k'} k' P(k') S(k',t_0) = \psi'(\theta(t))/\psi'(1)$.  So given that
$v$ is initially susceptible, the probability that $v$ is susceptible
at a later time $t$ is $\psi'(\theta(t))/\psi'(1)$.  Since the
probability $v$ is initially susceptible is $\phi_S(t_0)$, we conclude
\[
\phi_S(t) = \phi_S(t_0) \frac{\psi'(\theta)}{\psi'(1)}
\]
The rate of change of $\phi_S$ is simply $\phi_S(t_0) \dot{\theta}
\psi''(\theta)/\psi'(1)$.  It is straightforward to see that the
amount that goes from $\phi_S$ to $\phi_{I,1}$ is $\phi_S(t_0)
\beta_1\phi_{I,1}\psi''(\theta)/\psi'(1)$ and the amount going into
$\phi_{I,2}$ is $\beta_2 \phi_{I,2}\psi''(\theta)/\psi'(1)$.

So in figure~\ref{fig:theta}, we have expressions for the flux along
each edge except the edges into $\phi_R$.  These edges are
straightforward, because the recovery rate for disease $1$ is
$\gamma_1$ and the recovery rate for disease $2$ is $\gamma_2$.  So
the total flux from $\phi_{I,1}$ to $\phi_R$ is $\gamma_1\phi_{I,1}$
and the flux from $\phi_{I,2}$ is $\gamma_2 \phi_{I,2}$.  

Using the flows in figure~\ref{fig:theta}, we can arrive at a coupled
system for $\theta$, $\phi_{I,1}$ and $\phi_{I,2}$.  It is
\begin{align*}
\dot{\theta} &= -\beta_1 \phi_{I,1} - \beta_2 \phi_{I,2}\\
\dot{\phi}_{I,1} &= -(\beta_1 + \gamma_1) \phi_{I,1} + \beta_1
\phi_{I,1} \phi_S(t_0)\frac{\psi''(\theta)}{\psi'(1)}\\
\dot{\phi}_{I,2} &= -(\beta_2 + \gamma_2) \phi_{I,2} + \beta_2
\phi_{I,2} \phi_S(t_0)\frac{\psi''(\theta)}{\psi'(1)}
\end{align*}
with $\theta(t_0)=1$ and $\phi_S(t_0)$,  $\phi_{I,1}(t_0)$, and
$\phi_{I,2}(t_0)$ given by the initial state of the population.

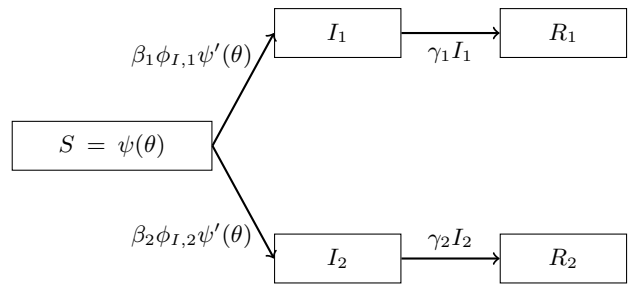
\begin{figure}
\begin{tikzpicture}
\node [wide] at (0,0) (S) {$S=\psi(\theta)$};
\node [int] at (3,1.5) (I1) {$I_1$};
\node [int] at (3,-1.5) (I2) {$I_2$};
\node [int] at (6,1.5) (R1) {$R_1$};
\node [int] at (6,-1.5) (R2) {$R_2$};
\path [->, thick,left,pos=0.8] (S.0) edge node{{$\beta_1\phi_{I,1}\psi'(\theta)$}} (I1.180);
\path [->, thick,left,pos=0.8] (S.0) edge node{{$\beta_2\phi_{I,2}\psi'(\theta)$}} (I2.180);
\path [->, thick,below] (I1) edge node{{$\gamma_1 I_1$}} (R1);
\path [->, thick,above] (I2) edge node{{$\gamma_2 I_2$}} (R2);
\end{tikzpicture}
\caption{Flow diagram leading to equations for the proportion of the
  population in each compartment.}
\label{fig:S}
\end{figure}

These equations govern the spread of the disease through the network.
However, they are not the usual variables of interest.  Typically we
want to know the proportion susceptible, infected, or recovered.
Figure~\ref{fig:S} shows a flow diagram governing the proportion of
the population in each state.  We can use this to recover $S$, $I_1$, $I_2$,
$R_1$, and $R_2$.  As we noted above, $S(t) =
\psi(\theta)$.  The flux into $I_1$ can be calculated to be
$\beta_1\phi_{I,1} \psi'(\theta)$, and the flux into $I_2$ is
$\beta_2\phi_{I,2}\psi'(\theta)$.  The fluxes from each of these into
the recovered states are $\gamma_1I_1$ and $\gamma_2 I_2$.  We
distinguish the two recovered states because we will frequently be
interested in the total proportion infected by each disease.  We could
have similarly subdivided $\phi_R$ into two compartments, but it would
not provide any information that is useful here.

Thus our final system of equations is
\begin{align*}
\dot{\theta} &= -\beta_1 \phi_{I,1} - \beta_2 \phi_{I,2}\\
\dot{\phi}_{I,1} &= -(\beta_1 + \gamma_1) \phi_{I,1} + \beta_1
\phi_{I,1} \phi_S(t_0)\frac{\psi''(\theta)}{\psi'(1)}\\
\dot{\phi}_{I,2} &= -(\beta_2 + \gamma_2) \phi_{I,2} + \beta_2
\phi_{I,2} \phi_S(t_0)\frac{\psi''(\theta)}{\psi'(1)}\\
S&=\psi(\theta)\\
\dot{I}_1 &= \beta_1\phi_{I,1} \psi'(\theta) - \gamma_1 I_1\\
\dot{I}_2 &= \beta_2\phi_{I,2} \psi'(\theta) - \gamma_2 I_2\\
\dot{R}_1 &= \gamma_1 I_1\\
\dot{R}_2 &= \gamma_2 I_2
\end{align*}

\paragraph{Regime Analysis}
There are several regimes that can be identified.  When the cumulative
number of infections is small enough that $S$ and $\phi_S$ are
approximately $1$, then we can neglect nonlinear terms.  Which regime
is observed is determined by the relative sizes of the two epidemics
when the linear approximation breaks down.  We focus our attention on
regimes for which the linear approximation is valid at the initial
time.  

When the initial condition is small, the linear terms dominate and the
epidemics grow (or decay) exponentially.  The equations governing the
epidemics are uncoupled in this regime.  Physically, this means that
competition for nodes is so weak that it can be neglected.  In fact a
stronger statement is true: in addition to not having inter-disease
competition, there is no intra-disease competition: a transmission
path is very unlikely to encounter a node infected along another
transmission path.

Assume that the exponential growth rates of the two diseases are $r_1$
and $r_2$, with $r_1 \geq r_2$.  This continues until one of them
becomes large enough that competition begins to appear.  At this point
the growth of both epidemics starts to slow.  If the difference in
epidemic sizes is large enough at this point, then this first epidemic
will not be slowed by the much smaller other epidemic.  The dynamics
will proceed as if there were just one disease spreading.  Eventually
the susceptible population will decrease, the disease will peak and
eventually decay away, all with the second disease negligible.  Once
this decay has occurred, there will be a ``residual'' network.  The
second disease will continue to spread along this network.  If the
residual network is well-enough connected (and the second disease
sufficiently infectious), the second disease can continue to grow, and
then it experiences its own epidemic.

If the diseases are close enough in size when nonlinear terms become
important, then both diseases contribute a non-negligible amount to
reduction in $S$ and $\phi_S$ at the same time.  Thus they interact
dynamically.  Each disease contributes in a non-negligible way to
hindering the spread of the other.  To determine whether this can
happen, we use a simple balance based on the initial sizes and the
early growth rates.   Typically we might expect
that one disease becomes large while the other is still exponentially small.

If one disease is sufficiently small when the other disease becomes
large, then the larger disease will spread and cause an epidemic that
is effectively the same size as it would be in the absence of the
smaller disease.  Using the initial sizes and growth rates, it is
straightforward to calculate the sizes of the two diseases once
nonlinearities begin to be significant.  For the two diseases to not
interact, the ``smaller'' disease must remain negligibly small until
the ``larger'' disease has finished its epidemic.  We derive slightly
different thresholds for the case where the smaller disease is the
fast-growing disease or the slow-growing disease.  To derive the
threshold condition, we make a crude assumption that the two diseases
continue spreading according to the linear growth rate.  In the
non-overlapping regimes, the smaller disease must remain small
throughout the spread of the larger disease.  We crudely choose to
apply our conditions when the exponential growth implies that the
larger disease would have reached size $1$.  We look at the size of
the smaller disease (assuming exponential growth) at this time.  For a
given observed size for the smaller disease, we reach different
conclusions if it is the fast or the slow growing disease.  If it is
the fast-growing disease, and we observe $0.05$, then that means that
at previous times it was much smaller, and so we would not expect to
observe any impact.  On the other hand if it is the slower-growing
disease, and we observe $0.05$, that means for much of the spread of
the larger disease it was at about that size, and so we would expect
to observe some impact.  So if the fast-growing disease is the small
disease, we allow it to be as large as $0.05$ when the slow disease
would reach $1$.  If the slow-growing disease is the small disease, we
require that it be much smaller, choosing $0.0025=0.05^2$ instead.
Our choice of threshold is somewhat arbitrary, and influenced by the
fact that $\ln 0.05 \approx -3$ giving a simple expression for our
thresholds.  Taking this, we can derive the $C_{\text{min}}\approx -6$
and $C_{\text{max}}\approx 3 r_2/r_1$ thresholds in the text.

Taking $\rho_1$ and $\rho_2$ to be the proportions initially infected
with each disease (at random).  So long as $C = \ln \rho_2 -
(r_2/r_1)\ln \rho_1$ is not between $-6$ and $3 r2/r1$, then the two
diseases will not interact dynamically.  One disease will become large
and run through its entire epidemic prior to the other disease
(possibly) having its own epidemic.  This defines the
``non-overlapping'' epidemic regimes.  If $C$ lies within this range,
then we have the overlapping epidemic regime.

\paragraph{Impact of Behavior change}

\begin{figure}
\begin{tikzpicture}
\node [wide] at (0,0) (phiS) {$\phi_S=\phi_S(t_0)\frac{\psi'(\theta)}{\psi'(1)}$};
\node [int] at (3,1.5) (phiD) {$\phi_D$};
\node [int] at (3,-1.5) (phiB) {$\phi_B$};
\node [int] at (6,1.5) (phiR) {$\phi_R$};
\node [int] at (6,-1.5) (1mT) {$1-\theta$};
\path[->, thick, left,pos=0.95] (phiS.0) edge node {{$\beta\phi_D\phi_S(t_0)\frac{\psi''(\theta)}{\psi'(1)}$}} (phiD.180);
\path[->, thick, left,pos=0.95, font = \footnotesize] (phiS.0) edge node {{$(\delta_D\phi_D+\delta_B\phi_B)\phi_S(t_0)\frac{\psi''(\theta)}{\psi'(1)}$}} (phiB.180);
\path[->, thick,above,sloped] (phiD.340) edge node {{$(\beta+\delta_D)\phi_D$}} (1mT.90);
\path[->, thick, above] (phiB) edge node {{$\delta_B\phi_B$}} (1mT);
\path[->, thick,above] (phiD) edge node {{$\gamma\phi_D$}} (phiR);
%\path[->, thick, right] (phiI2) edge node {{$\gamma_2\phi_{I,2}$}} (phiR);
\end{tikzpicture}\\[20pt]
\begin{tikzpicture}
\node [wide] at (0,0) (S) {$S=\psi(\theta)$};
\node [int] at (3,1.5) (I) {$I$};
\node [int] at (3,-1.5) (B) {$B$};
\node [int] at (6,1.5) (R) {$R$};
\path [->, thick,left,pos=0.8] (S.0) edge node{{$\beta\phi_D\psi'(\theta)$}} (I.180);
\path [->, thick,left,pos=0.8] (S.0) edge node{{$(\delta_D\phi_D+\delta_B\phi_B)\psi'(\theta)$}} (B.180);
\path [->, thick,below] (I) edge node{{$\gamma I$}} (R);
\end{tikzpicture}
\caption{The flow diagrams underlying the spread of an infectious
  disease in the presence of a behavior change.  We assume behavior
change can be triggered by contact with an infected individual or with
an individual who has already adopted the change.  We assume behavior
change provides complete immunity to disease.}
\label{fig:behave}
\end{figure}
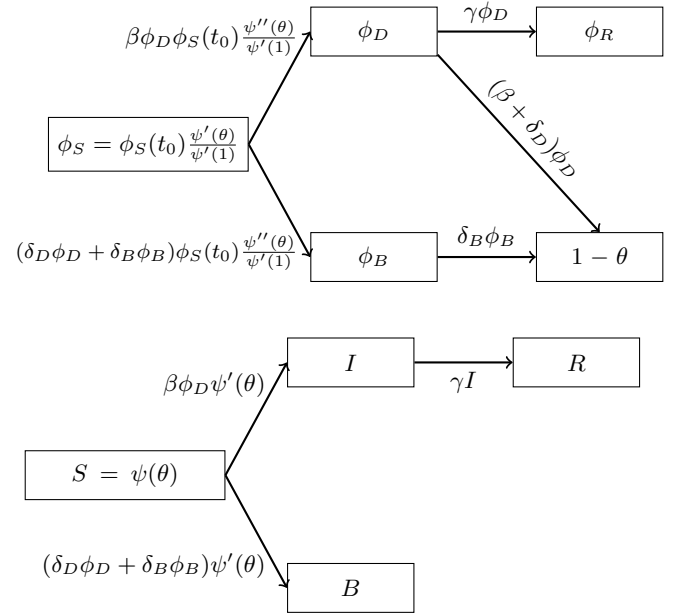

We now consider the spread of a single disese through a network, which
can be prevented with a behavior change.  We assume that the behavior
change gives complete protection from infection.  Once an individual
has adopted the behavior change, the change is permanent.  An
individual who has changed behavior will transmit that behavior change
to partners.

The disease transmits at rate $\beta$ and recovery occurs at rate
$\gamma$.  Contact with an infected individual causes behavior change
at rate $\delta_D$.  Contact with an individual whose behavior has
changed causes behavior change at rate $\delta_B$.  The flow diagrams
are shown in figure~\ref{fig:behave}.  The resulting equations are
\begin{align*}
\dot{\theta} &= -(\beta_D+\delta_D)\phi_D - \delta_B\phi_B\\
\dot{\phi}_B &= (\delta_B\phi_B + \delta_D\phi_D)\phi_S(t_0)\frac{\psi''(\theta)}{\psi'(1)}\\
\dot{\phi}_D &= \beta_D\phi_D\phi_S(t_0)
\frac{\psi''(\theta)}{\psi'(1)} - (\beta_D +\delta_D + \gamma)\phi_D
\\
\dot{I} &= \beta_D \phi_D \psi'(\theta) - \gamma I\\
\dot{B} &= (\delta_B\phi_B+\delta_D\phi_D) \psi'(\theta)\\
\dot{R} &= \gamma I\\
S &= \psi(\theta)
\end{align*}
The early growth of $\phi_D$ is exponential with rate
$\beta_D\phi_S(t_0)\psi''(1)/\psi'(1) - (\beta_D+\delta_D+\gamma)$.
If $\delta_D$ is sufficiently large, the infection decays.  This
corresponds to a balance between transmitting disease prior to either
recovering or transmitting behavior change.  When the transmission
probability is small enough a typical infected individual causes fewer
than $1$ new infection and the disease must die out.

Changing $\delta_B$ does not alter the early growth rate of the
disease.  So we might anticipate that the disease will be able to
cause a large scale epidemic.  However, if $\delta_B>0$, the behavior
change itself also leads to an ``epidemic''.  If the behavior growth
rate is sufficiently large, its ``epidemic'' occurs while the disease
epidemic is still exponentially small.  In this limit, the behavior
change will dominate the populaion, and the disease remains
exponentially small.  We do not see the complementary case in which
the disease becomes large while the behavior change is exponentially
small, because as disease incidence increases it directly induces
behavior changes.  So either the behavior change has its ``epidemic''
first, or the two have overlapping ``epidemics''.

\bibliographystyle{plain}
\providecommand{\noopsort}[1]{}

\end{document}